# Uncertainty-Informed Screening for Safer Solvents Used in the Synthesis of Perovskite via Language Models

Arpan Mukherjee[1], Deepesh Giri[2], Krishna Rajan[1*]

[1]Department of Materials Design and Innovation, University at Buffalo, Buffalo, NY, 14260 – 1660, USA

[2]Long Island University, Brooklyn, NY 11201, USA

## Abstract

The challenge of accurately predicting toxicity of industrial solvents used in perovskite synthesis is a necessary undertaking but is limited by a lack of a targeted and structured toxicity data. This paper presents a novel framework that combines an automated data extraction using language models, and an uncertainty-informed prediction model to fill data gaps and improve prediction confidence. First, we have utilized and compared two approaches to automatically extract relevant data from a corpus of scientific literature on solvents used in perovskite synthesis: smaller bidirectional language models like BERT and ELMo are used for their repeatability and deterministic outputs, while autoregressive large language model (LLM) such as GPT-3.5 is used to leverage its larger training corpus and better response generation. Our novel '*prompting and verification'* technique integrated with an LLM aims at targeted extraction and refinement, thereby reducing *hallucination* and improving the quality of the extracted data using the LLM. Next, the extracted data is fed into our pre-trained multi-task binary classification deep learning to predict the ED nature of extracted solvents. We have used a Shannon entropy-based uncertainty quantification utilizing the class probabilities obtained from the classification model to quantify uncertainty and identify data gaps in our predictions. This approach leads to the curation of a structured dataset for solvents used in perovskite synthesis and their uncertainty-informed virtual toxicity assessment. Additionally, chord diagrams have been used to visualize solvent interactions and prioritize those with potential hazards, revealing that 70% of the solvent interactions were primarily associated with two specific perovskites.

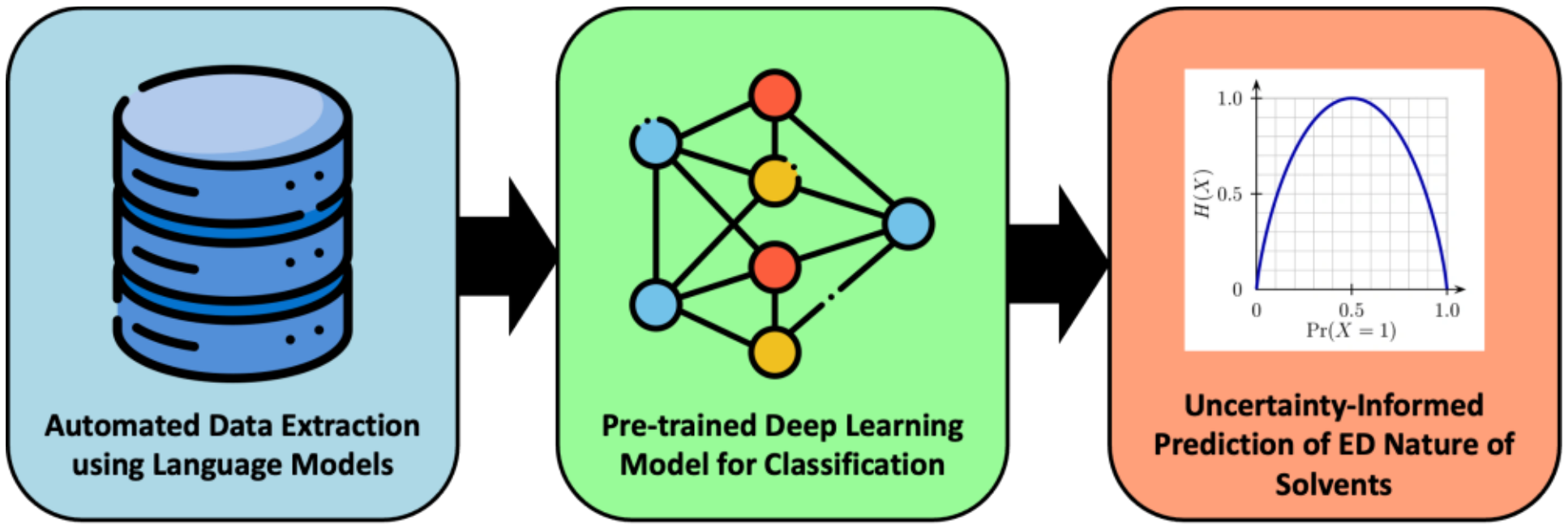

*Graphical Abstract of the proposed framework for identifying the endocrine-disrupting (ED) potential of solvents used in perovskite synthesis. The first step is automated data extraction using language models. The second step uses a pre-trained deep learning model to classify the extracted solvent data. The third step shows the uncertainty-informed prediction of the ED nature of solvents, incorporating Shannon entropy-based uncertainty quantification.*

* Email: krajan3@buffalo.edu

## 1. Introduction

Perovskites are gaining importance as the most promising photovoltaic materials due to their low production costs and high photoconversion efficiencies, which now exceed 20%[1–4]. These materials are manufactured by both solution-based[5] and solid-state techniques[6]. The more common solution-based methods utilize organic solvents that significantly influence film formation, reaction rate, and overall quality. There is abundant literature on the synthesis of perovskite solar cells[2,7–9], the usage of solvents in such synthesis[5,10–12], and solvent selection guides [13–16]. However, structured datasets on the solvents used in perovskite synthesis and their accurate toxicity assessments are lacking, raising safety and environmental hazard issues. Existing datasets on solvents used for Perovskite synthesis mainly cover a limited set of solvents, such as DMF, DMSO, GBL, and IPA[17]. Given the scarcity of comprehensive data, the necessity to extract comprehensive information from scientific literature is crucial. Automated Data Extraction using language models has gained popularity as a tool for extracting tailored data in materials science[18–20]. In this study, we have developed two methods for hierarchical knowledge extraction using language models to systematically compile and verify information on solvents used in perovskite synthesis (see Figure 1). We have designed a prompting technique that incorporates a feedback mechanism to iteratively refine the accuracy of data extraction using language models.

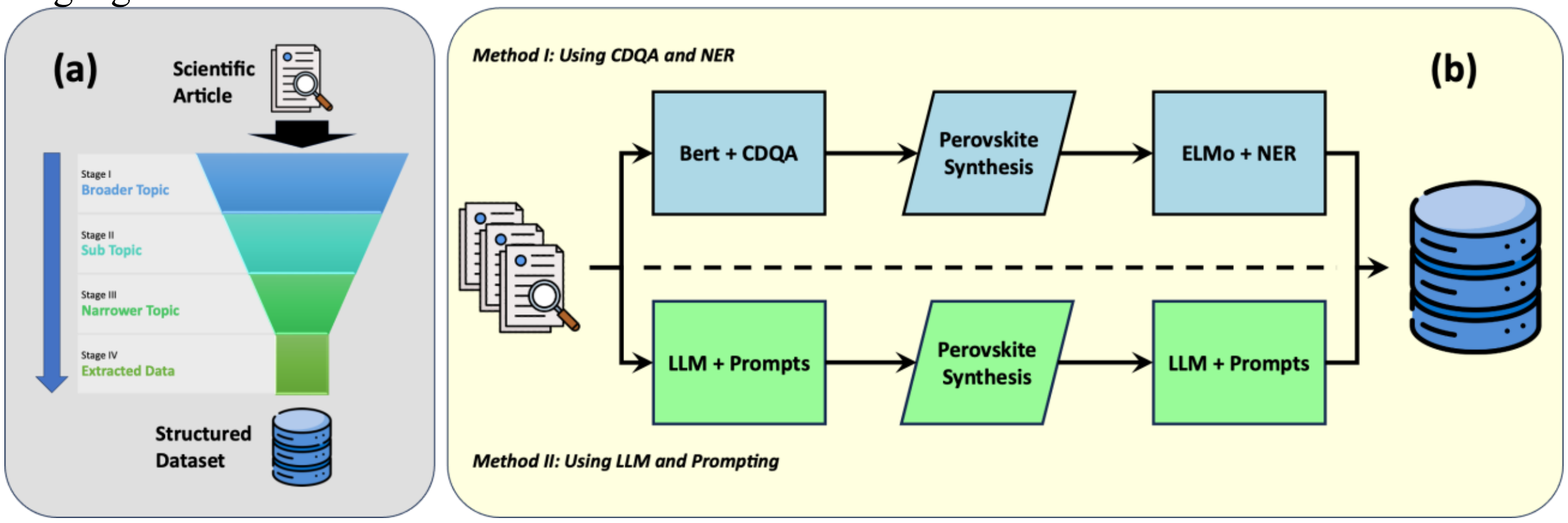

*Figure 1: Automated Data Extraction and Curation using Language models. (a) Two different methods for implementing the hierarchical extraction process. Method 1 uses a combination of CDQA and NER to extract and refine information. Method II employs LLMs with prompting to achieve the same objective, showcasing different approaches to achieve accurate data extraction from research articles. (b) Hierarchical extraction process, where information is progressively refined from a broad topic to narrower, more specific details. This involves a series of stages, starting with a general context and moving through subtopics to ultimately extract precise data.*

A hierarchical knowledge extraction methodology using language models is implemented that progresses from broad to narrow topics. (see Figure 1(a)). This approach ensures a comprehensive extraction of relevant information while maintaining contextual accuracy and precision and is, hence, well suited for sparse data[21]. Method I has a straightforward sequence involving the use of a contextual model and a combination of Closed Document Question Answering (CDQA) and Named Entity Recognition (NER). Early contextual language models such as ELMo[22], BERT[23,] and GPT-2[24] have significantly improved understanding the sequence-level semantics and have shown state-of-the-art performance in several NLP tasks such as sequence classification[25], question answering[26,27], language modeling[28] and translation[29,30] requiring fewer parameters and training time. Other NLP techniques, such as Closed Document Question Answering (CDQA) and Named Entity Recognition (NER), benefit from these advances, as data extraction has seen higher efficiency and accuracy (see Figure 1(b).). However, the reliance on specific contextual models integrated with CDQA and NER to identify chemical entities such as solvents presents challenges, primarily due to the scarcity of high-quality, chemically focused training data. This scarcity often results in a higher likelihood of type I errors (false positives) compared to type II errors (false negatives).

Method II uses more recent Large Language Models, such as GPT 3.5, along with designed prompts for the hierarchical automated data extraction. LLMs have brought new capabilities that differ from earlier contextual models by utilizing a high number of self-attention layers and a more extensive training corpus. These features enable them to generate more accurate and diverse responses and better generalize across various tasks without the explicit need for task-specific downstream architectures like CDQA and NER. As shown in Figure 1(b), prompt engineering becomes essential when utilizing the in-built response generation capabilities of LLMs, as it replaces the role of traditional NLP tools by allowing the model to adapt its responses based on finely tuned prompts[19]. This method leverages the built-in response generation capabilities of the LLMs, enabling the identification and classification of chemical entities such as solvents directly through well-designed prompts rather than integrating them with separate tools. Furthermore, the use of domain knowledge is essential for designing and refining the prompts to evaluate the relevance and accuracy of the LLM's responses. During inference, LLMs process text at the token level, predicting the next token in a sequence given the preceding tokens. This capability allows them to assign probabilities to different tokens, including those corresponding to named entities like solvents, based on the context provided. Thus, LLMs are capable of performing both CDQA and NER tasks through their all-purpose design, eliminating the need for additional specialized tools. The contrast between the two methods for automated data extraction is shown in Figure 1(b).

However, LLMs suffer from a phenomenon called hallucination, where models assert the truth of a statement based on its resemblance to training data, regardless of its actual logical or factual basis [31]. The models use named entities as "indices" to access memorized data, leading to false positives when these entities are recognized from the training set[31]. Moreover, LLMs identify solvents and other specific entities based on the context provided in the query and its training corpora rather than acting as a classification model. Hence, LLMs often "fill in with common knowledge" during question answering due to “overgeneralization” and “overgeneration,” reflecting biases and prevalent information from their training data[32]. LLMs struggle to generate accurate information for complex or less frequent queries due to gaps in training data or the complexity of the task[32]. Thus, prompts must be carefully designed and layered, incorporating human feedback and domain knowledge to sequentially target the information towards the specific context. We have developed a novel *prompting and verification* technique that targets particular data to be extracted and performs a self-check to mitigate hallucination. At the broader level (as per Figure 1(a)), initial prompts are designed to gather general information and context about the broader subject. At the subtopic level, follow-up prompts are utilized to delve into specific areas identified from the broader context, extracting relevant information that outlines a specific topic of interest within the larger topic. As we move down to the narrower topic level, the method of extraction becomes more refined, targeting specific information within each subtopic to gather data points relevant to the focused areas.

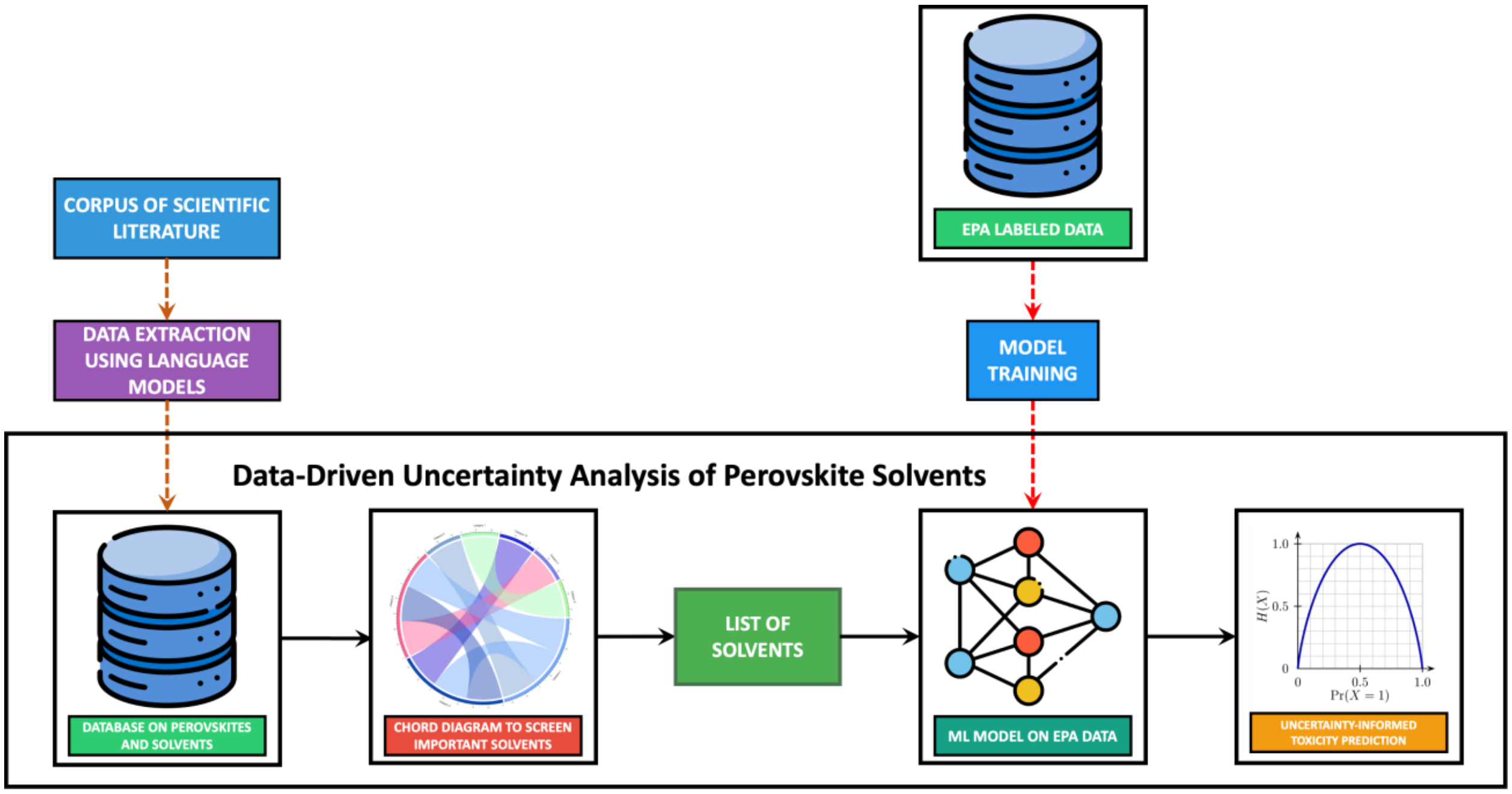

*Figure 2: Workflow for Uncertainty-Based Classification of Solvents Using Machine Learning. The pipeline starts with the collection of scientific literature, followed by the extraction of data using language models. A chord diagram is utilized to screen and prioritize important solvents. This information, along with EPA-labeled data, is used to train a machine learning model. The workflow culminates in the application of the trained model on EPA data to achieve uncertainty-informed classification.*

The key innovation of this work is the integration of uncertainty quantification into the classification model towards toxicity prediction for the targeted application, where the data for the application is extracted using language models. The solvents used in perovskite synthesis could pose a harmful risk to occupational health, safety, and the environment[33]. Hence, it is essential to identify toxic chemicals with the aim of eliminating or reducing their use and developing safer alternatives. Figure 2 illustrates the framework for uncertainty-informed model prediction following the data extraction using language models. We first use a chord diagram to visualize and prioritize solvents based on their interactions and potential hazards. Next, we have utilized an uncertainty-informed approach to assess the endocrine-disrupting effects of identified solvents by calculating the Shannon entropy on the probability of outcomes from a recently developed deep neural network[34]. The quality of prediction for toxicity assessment has been measured in recent research using variability in ensemble learning and latent space distance metric[35]. Under these methods, the uncertainty associated with each data point is computed with respect to other training samples, and thus, discrepancies that may arise due to data from different sources are not taken into account. Model prediction uncertainty for classification algorithms is computed using the two most commonly used methods that rely on the training phase of the ML model. The first method, deep ensemble,[36,37,] uses an ensemble of deep learning models with fixed model architecture and different random initial weights. The uncertainty in prediction is obtained from the point estimates of model prediction for the different models. The second method, Monte Carlo dropouts,[38,39] assigns random values of dropouts during the training phase of the network. The uncertainty in model prediction is calculated in the same way as the deep ensemble method using the point estimates from the different trained models. A third less common method can be used whereby the model prediction for a chemical is compared with the nearest chemicals in the training dataset in the embedded space of the penultimate layer of the neural network[40]. However, once the model is trained, a compelling method for evaluating variability in prediction is conditioned on the fixed model structure and the learned parameters. We have adopted a Shannon entropy-based uncertainty quantification (UQ) method to quantify the uncertainty in prediction that may arise due to different sources of data. It is agnostic to the type of model architecture without further requiring any additional training. Shannon entropy is a well-known measure of uncertainty[41] and finds its application in classification, parameter learning, and active learning[42].

Our proposed framework for uncertainty-informed predictions bridges the gap between sparse data buried in scientific literature and real-world applications, addressing the safety and sustainability of perovskite synthesis. The rest of the paper is organized as follows: Section 2 details the implementation of the two methods for automated data extraction using language models and uncertainty-informed classification using deep learning. Section 3 first visually explains the distribution of keywords in the literature on perovskite synthesis based on our data extraction method. We then use a deep learning model and Shannon entropy to make uncertainty-informed toxicity predictions for selected solvents. Finally, Section 4 concludes the paper.

## 2. Methodology

### 2.1 Dataset

We have downloaded 2000 peer-reviewed articles providing 30,000 paragraphs that serve as metadata for information retrieval. The DOIs for the articles were queried by searching for the phrases – “halide perovskites,” “hybrid organic, inorganic perovskites,” “toxic perovskites,” “perovskite solar cells,” and “chemical synthesis of perovskites” on CrossRef[43]. Following this, the articles were acquired from open-access journals such as Nature, American Chemical Society, Elsevier, and Royal Society of Chemistry. These articles form the metadata on which we implement contextual NLP to get data for further analysis.

### 2.2 Method I: Early Contextual Models

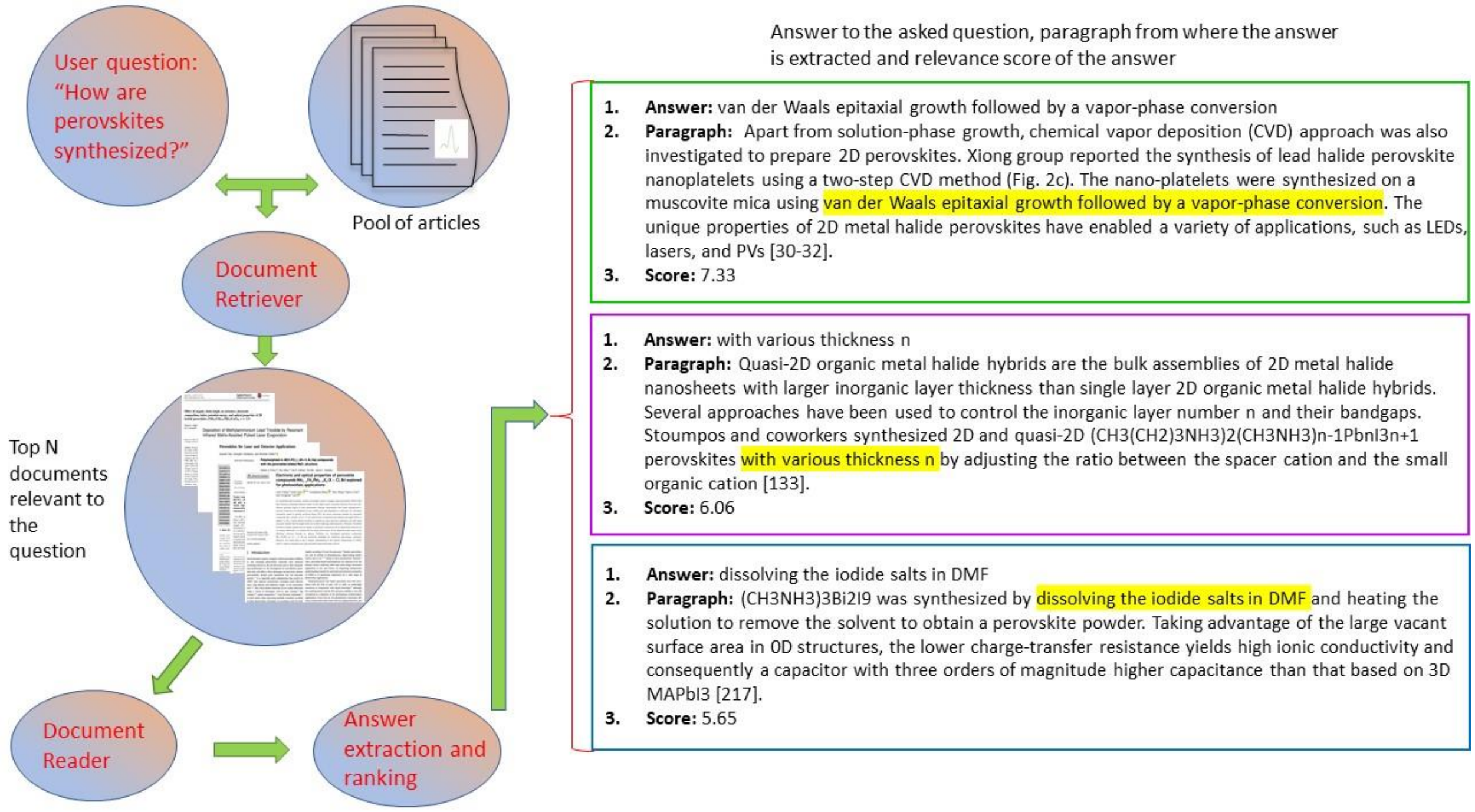

*Figure 3: In general, the question-answering (QA) system in NLP can be divided into two categories – Open Domain Question Answering (ODQA) and Closed Domain Question Answering (CDQA). The ODQA is capable of answering questions from any field, while the CDQA answers questions only from a specific domain of knowledge. Google Assistant, Amazon Alexa, etc., are examples of the ODQA, while chatbots are examples of the Closed Domain systems. In this work, we use CDQA to identify the relevant paragraphs on perovskite synthesis that serve as metadata for further analysis. The 'Document Retriever' scans the given pool of articles to filter out the 'N' most relevant documents to the given question. The 'Document Reader' processes these documents to get the closest possible answers. In this work, we extracted three answers from each article. We also acquired the corresponding paragraphs where the answers are based and used them to get the perovskites and the solvents. Answers with higher scores appear more relevant than the others.*

As per Figure 1, we have integrated BERT as a language model with Closed Document Question Answering (CDQA) followed by ELMo with Named Entity Recognition (NER) to automate the data extraction process[44], enabling the hierarchical knowledge extraction from a broader topic to a structured dataset. Figure 3 explains how the CDQA works. CDQA is an NLP subtask that involves asking context-specific questions within a closed domain, such as perovskite synthesis, extracting relevant paragraphs or sentences from a scientific article without having to manually annotate them. There are two main components of the CDQA system – Document Retriever and Document Reader. The Document Retriever identifies a list of 'Top N' candidate documents that are likeliest to the context of perovskite synthesis using similarity metrics. We have used cosine-similarity between the TF-IDF features of the documents and the phrase "perovskite synthesis." Next, these documents are divided into paragraphs and fed to the Document Reader, BERT, which gives the most probable paragraphs to the question "How are perovskite synthesized." The answers were compared and ranked in the order of the model score, which is given by the softmax probability derived from the last layer of the BERT model. At the end of this step, three paragraphs most relevant to perovskite synthesis are extracted from each 'Top N' candidate document.

NER is the second subtask of our NLP pipeline that classifies keywords extracted from a given paragraph. Commonly available NER tools are ChemicalTagger[45], OSCAR4[46], Chemical Named

Entities Recognition[47], and ChemDataExtractor[48], each trained for identifying specific terminologies and contexts within the materials science domain. In this work, to extract all the chemicals (perovskites, solvents, etc.), we used an ELMo-based NER tool developed by Kim et al.[49]. The NER model developed by Kim et al.[49] uses a classification model that is trained on an internal database of over 2.5 million materials science articles. The details of the architecture training accuracy are given in the Supporting Information. At the end of this step, a structured dataset is formed by listing perovskites and their corresponding solvents that can be used for downstream tasks such as toxicity prediction.

A critical limitation of Method I is that the segmentation is typically conducted at the paragraph level rather than considering token-level constraints. This approach can overlook nuanced details that may span multiple sentences or paragraphs within a single paper. Crucial information about the interaction of solvents with perovskite materials might be dispersed across several sentences or paragraphs within a single research paper, but the paragraph-level segmentation used in CDQA could overlook these interconnected details. The CDQA method often treats each paragraph as an isolated unit during information retrieval, potentially missing valuable connections that could exist across different sections or even pages of the same document. This fragmented approach can lead to information loss, similar to the challenges encountered in Retrieval-Augmented Generation (RAG) models, which also struggle with integrating information across fragmented document sections. Furthermore, as pointed out before, the solvents identified by the NER model may be restricted to entities present in its training dataset, highlighting the necessity for a context-based approach to accurately identify solvents beyond the dataset's limitations.

### 2.3 Method II: LLM and Prompting

Generative models, like GPT 3.5, trained on vast corpora, have a broader knowledge base that enables them to synthesize answers by integrating information across entire texts and thereby establish connections between prompts and specific scientific concepts like perovskites, which is beyond the capability of Method I. As explained earlier, the hierarchical information extraction using LLM requires careful design of prompts. We first explain the method of using prompts and LLMs for a particular level by detailing the steps involved in extracting and verifying information from research articles (see Figure 4).

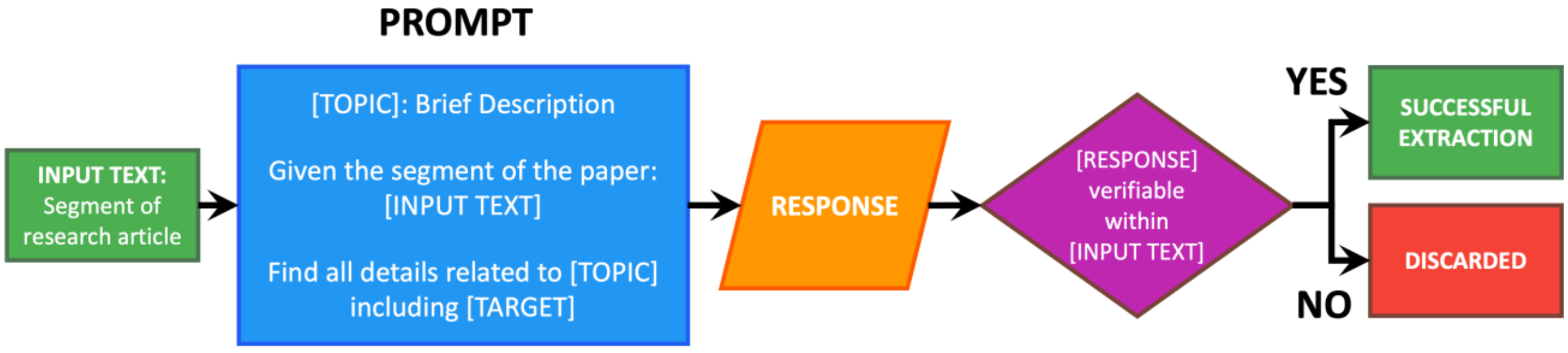

*Figure 4: Flowchart for information extraction from a research article using prompting and verification technique, starting with the "Input Text" box where the paper segment is specified, followed by a "Prompt" box detailing the search query. The process then moves to a "Response" diamond indicating LLM response, which leads to either "Successful Extraction" or "Discarded" based on the verifiability within the input text.*

Figure 4 shows that we employ a structured *prompting and verification* process to extract and verify specific information from a predefined segment of a research article. Responses from Method I, which provides the most relevant paragraphs, are used to design prompts through a trial-and-error process. OpenAI Playground[†] offers an interactive dashboard to experiment with various models and parameters, allowing users to fine-tune and test prompts in real time. Given an input text segment, a prompt is generated to find all details related to the topic. While extracting information from a text segment on a specific [TOPIC], the LLM is prompted with a brief [DESCRIPTION] of the [TOPIC] along with the text segment [INPUT TEXT]. The [TARGET] denotes the type of information to be extracted from a given segment. This differentiates our approach from traditional prompting by explicitly contextualizing the query within the prompt, ensuring that the LLM search is focused and relevant to the specific topic[18]. Since scientific texts often contain complex syntactic structures, nested entities, and domain-specific terminologies, it is important to include details related to questions in the prompt to extract the correct information[50]. This step is followed by a verification through subsequent prompting[20], where the LLM checks if the response details from the previous prompt are explicitly found within the provided input text segment. This strategy helps mitigate hallucinations by increasing specificity until the LLM produces the correct answer that is guided by accurate responses known from previous steps.

The prompting and verification technique is applied iteratively at each level, progressively narrowing down from broad topics to specific details by refining prompts and verifying responses (see Figure 5). Too many promptings can be cost-intensive; thus, care is given so that the target dataset can be obtained without excessive prompting. At each layer, the text from the previous layer is segmented based on the token limit of the LLM model. This segmentation approach utilizes the analytical capabilities of the LLM to interpret complex scientific data by concentrating on a smaller window for contextual understanding. The responses from multiple segments of a single paper are then consolidated using the LLM to form a coherent and comprehensive summary, which streamlines the relevant sparse and disparate information into an easily accessible form. The [TOPIC]s and their brief [DESCRIPTION]s for each layer are given in Table 1. Domain expertise, along with trial-n-error and the responses from Method I, have been used to come up with the descriptions. The first TOPIC is 'Perovskite,' where the description is targeted to establish a foundational understanding of the material.

[†] https://platform.openai.com/playground/chat?models=gpt-3.5-turbo

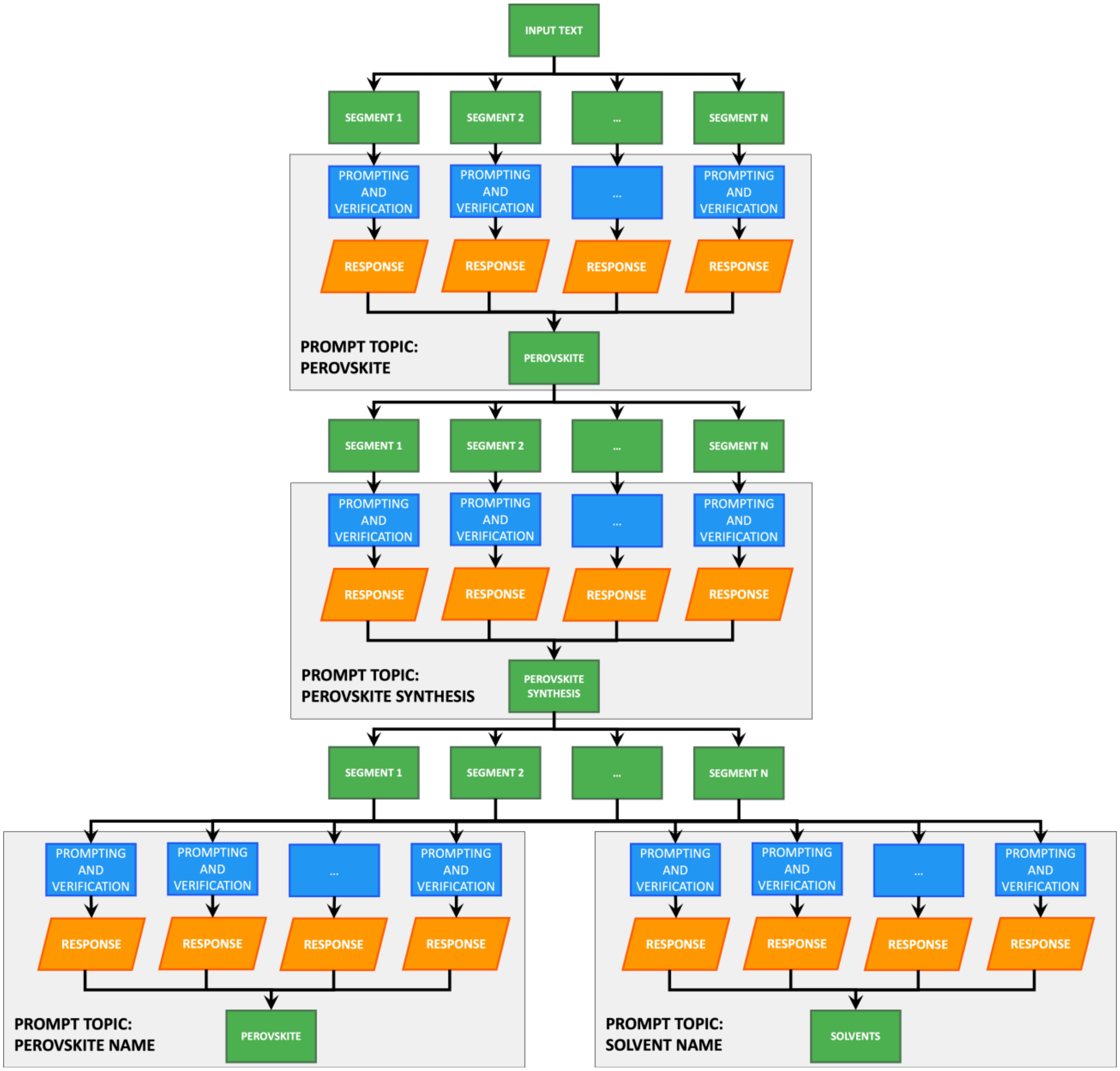

*Figure 5: Iterative hierarchical knowledge extraction process using LLMs. The input text is segmented into smaller chunks, each undergoing prompting and verification to extract responses relevant to the broad topic (Perovskite). These responses are then combined and re-segmented for the next level of specificity (Perovskite Synthesis), where the process is repeated. Finally, the combined responses are further segmented and processed at the narrowest levels, which include both Perovskite Name and Associated Solvent, ensuring accurate and detailed extraction of specific information.*

*Table 1: TOPICS and their brief descriptions used for prompting and extraction of data using layer-wise prompting and verification process shown in Figure 4.*

| TOPIC | Description | Targeted Information |
|---|---|---|
| Level 1: Perovskite | Perovskite has a unique crystal structure with the formula ABX3, where 'A' and 'B' are cations and 'X' is an anion, forming a three-dimensional network that contributes to the unique properties of perovskites, such as their excellent electronic and ionic conductivity. | Perovskites, including their chemical compositions, synthesis processes, and |
| | | various applications |
| Level 2: Perovskite Synthesis | Perovskite synthesis involves steps such as precursor preparation, dissolution in solvents, deposition, and subsequent annealing and crystallization to form the ABX3 crystal structure. | Chemistries related to perovskite synthesis, such as precursor, perovskite, and solvents |
| Level 3: Perovskite Name | Specific form of the ABX3 crystal, where ‘A’ and ‘B’ are cations and ‘X’ is an anion. | Name of the Perovskite Crystal in ABX3 Form |
| Level 3: Solvent Name | Solvents in perovskite synthesis are organic chemicals used to dissolve the precursors | Name of the Organic Solvent |

The second [TOPIC] is ‘Perovskite Synthesis,’ aimed at understanding the processes involved in creating perovskites. The prompt at this level extracts detailed information about the synthesis steps, including precursor preparation, dissolution in solvents, deposition, and subsequent annealing and crystallization. The responses from Level 2 are manually compared against the responses from the CDQA in Method I to check for the correctness of the prompting method. The third level focuses on more specific details, divided into two subtopics: ‘Perovskite Name’ and ‘Solvent Name.’ This step is similar to the NER step of the previous method, where instead of using a classification model, we rely on the LLM’s inherent understanding of context and scientific terms. The ‘Perovskite Name’ prompt seeks to identify specific forms of the ABX3 crystal by listing the various cations and anions that define different perovskite compounds. It is to be noted that at any level, there can be multiple subdivisions based on the specific information needed, where subdivisions refer to narrower topics or categories derived from the broader topic to extract detailed and relevant data. The ‘Solvent Name’ prompt extracts information on the organic chemicals used in the synthesis process to dissolve precursors. The division into 'Perovskite Name' and 'Solvent Name' has been deliberately done to ensure that the LLM can accurately identify the named entities by using separate prompts and descriptions for each. Additionally, as explained earlier, the larger training corpus for GPT 3.5 eliminates the need for a separate NER component for identifying perovskites and solvents. The description of the terms added to the prompts aids in identifying the context of these terms better, while the [TARGET] targets the LLM toward specific data to be extracted. Furthermore, the hierarchical extraction allows data to be extracted at each level, and the data from each level can be repurposed for other research objectives, such as identifying precursor materials from the 'Level 2: Perovskite Synthesis' responses or evaluating device performance from the 'Level 1: Perovskite' responses.

## 2.4 Uncertainty-Informed Classification

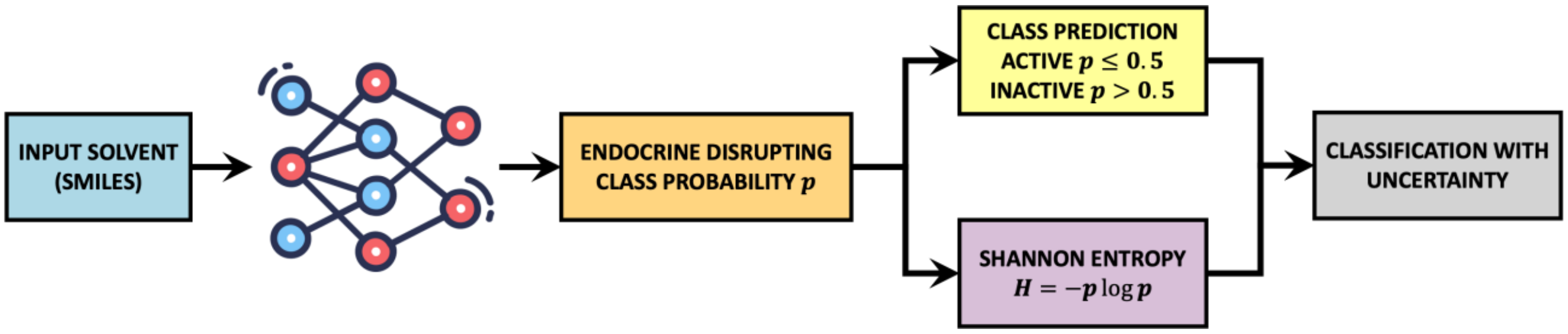

*Figure 6: Workflow for assessing the prediction uncertainty of endocrine disrupting chemicals. The process begins with the input of solvent data in the form of SMILES codes, which are processed by a deep neural network model to generate the class probability p of a solvent being endocrine disrupting. This probability is then used for class prediction (active/inactive) and for calculating Shannon entropy $H = -p \log p$ to assess the uncertainty of the classification. The final output is the classification with an associated uncertainty measure.*

In this section, we assess the prediction associated with determining if a given solvent is an endocrine disrupting chemical to underpin its decision-making and improve confidence in the prediction. This uncertainty in prediction is assessed by calculating the Shannon entropy from the class probabilities generated by the deep learning model used for classifying the solvents, which classifies each solvent as either active or inactive based on its potential endocrine-disrupting nature (see Figure 6). The ED nature of these solvents is assessed in terms of two classes of endocrine disruption: the 'Agonist' and 'Binding' class. These classes denote a molecule's potential interaction with the estrogen receptor. The deep neural network classification model used here is a stack of ten convolutions and two LSTM layers, followed by two dense layers. The final layer has two diverging sigmoid layers that output the probabilities for each class: 'Agonist' and 'Binding.' The model is trained on the SMILES (Simplified Molecular-Input Line-Entry System) codes[51] of organic chemicals obtained from the ToxCast[52] and Tox21[53] databases. SMILES are machine-readable notations that are used to represent chemical species in a single line using a set of ASCII characters. The convolution layers progressively extract the spatially correlated local features from the SMILES, while the LSTM layers are used for sequential data processing. Details of model accuracies are reported in the Supporting Information.

Shannon entropy, using the class probabilities provided by the sigmoid layers, provides a post-prediction uncertainty analysis[37,54] that assesses the precision of the data-driven model by quantifying the uncertainty associated with the predictions. Uncertainty in prediction can arise from different sources. The ML model,[34] trained on the list of EDCs from the ToxCast and Tox21, needs to be representative of organic molecules in general to obtain an interpretable prediction to accurately classify a solvent as either active or inactive for each class. Data from different sources may not necessarily follow the same probability distribution. Furthermore, there are aleatory uncertainties since we are relying on statistical methods to obtain the list of chemicals used for perovskite synthesis. Thus, there is a need to acknowledge the problem of data belonging to different distributions using relevant UQ methods. UQ converts the point prediction into a probabilistic prediction to gain more confidence in our prediction.

The prediction probability density function (or mass function for discrete output) conditioned on the model structure is given as:

$$p_i = p(y_i) = p_F(y_i|\boldsymbol{x}, D) \qquad (1)$$

The class probability using the last sigmoid layer of the deep learning model given in Figure 6 can be written as:

$$y_i = \sigma_i\big(F(x)\big) \qquad\qquad i = 1,2$$
$$\sigma_i = \frac{1}{1 + e^{-\beta_i F(x)}} \qquad (2)$$

Where $F(x)$ represents the input to the sigmoid function from the preceding layers of the neural network. This function maps the input features of a solvent to a probability $p_i$ indicating the likelihood of the solvent being an EDC. Also, $i = 1, 2$ determines the class of EDC (Agonist or Binding) and $\sigma$ is the sigmoid function. Given an organic molecule $x_j, j = 1$ to $N$ belonging to the list of solvents given in Table 2, the prediction probabilities $p_{ij}$ are given by the function $p_{ij} = \sigma_i\left(F(x_j)\right), p_{ij} \in [0,1]$. The relationship between uncertainty and output probability is not linear. The classification model can have low activation values in all the remaining neurons but still can have high sigmoid values. Thus, using only the sigmoid output as a measure of model uncertainty can be misleading. Shannon entropy removes this drawback by weighing the prediction probability $p_{ij}$ with the logarithm of the reciprocal of $p_{ij}$ and thereby used to measure the information content of each prediction. The basic intuition behind such formulation is that the unlikely event will be more informative, while the likely events have little information, and the extreme case events should have no information. The self-information or Shannon information function is the information content associated with a single prediction and is defined as:

$$I(p_i) = -\log p_i \qquad (3)$$

The Shannon entropy for the $j^{th}$ solvent for the $i^{th}$ class is measured as:

$$H_{ij} = -p_{ij}\log p_{ij} - (1 - p_{ij})\log(1 - p_{ij}) \qquad (4)$$

This calculation effectively captures the uncertainty of the prediction by considering both the probability of the event occurring and not occurring. This measure reaches its maximum when $p = 0.5$, indicating maximum uncertainty, and is minimal (zero) when $p$ is 0 or 1. The maximum entropy or the total uncertainty for the whole list of solvents for $j^{th}$class (Agonist or Binding) is $S_j = \sum H_{ij}$. The uncertainty associated with each $i^{th}$ solvent for the $j^{th}$ class of EDC is estimated as the ratio of the prediction entropy $H_{ij}$ and the maximum entropy $S_j$, providing a normalized measure of the uncertainty across all solvents in a class.

## 3. Results

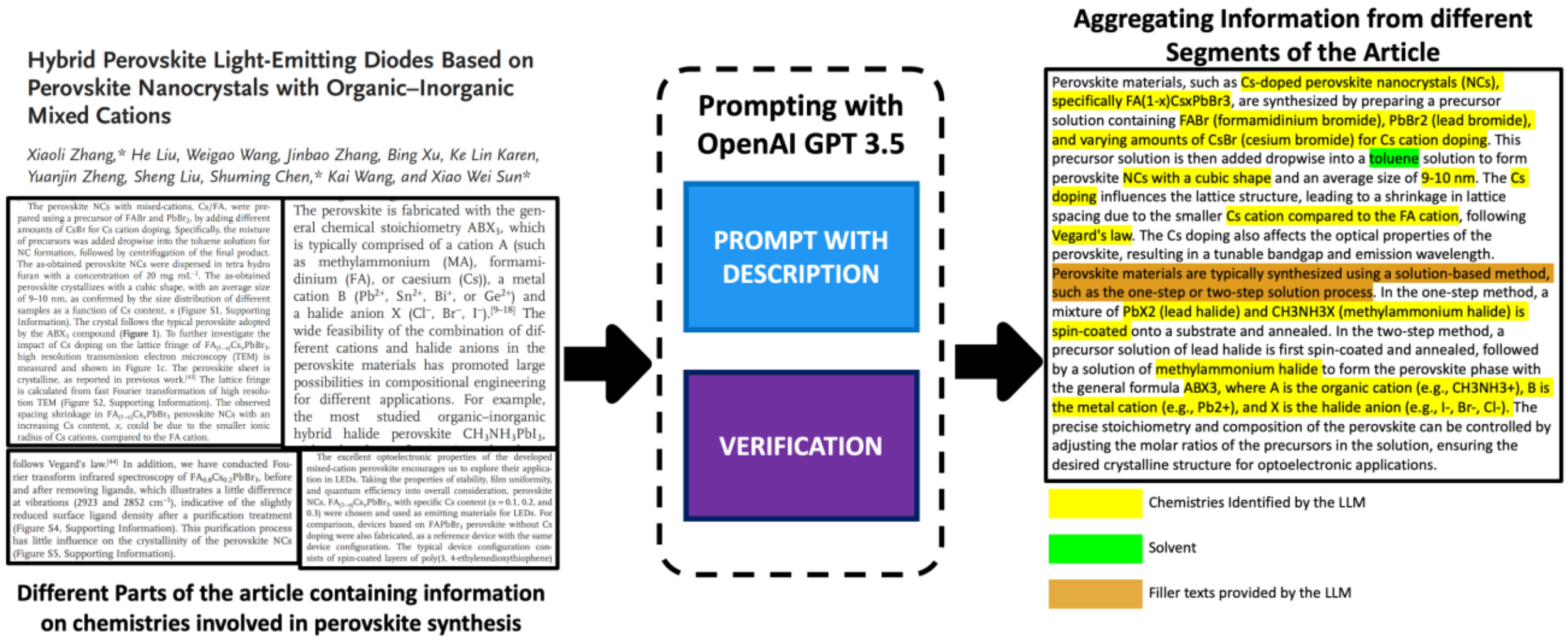

*Figure 7: Data Extraction using Method II demonstrating the ability of our method to fuse information from different sections of a research paper to extract detailed chemical information related to perovskite synthesis. The highlighted sections show various mentions of solvents, cations, and synthesis methods scattered throughout the document. The Method II method successfully integrates these disparate pieces of information. Results from Method I and II are reported in the Supporting Information.*

In this work, we have identified 35 different solvents using Method I and 54 solvents using Method II that are used during perovskite synthesis. The Supporting Information contains a table of all solvents identified by the two methods. A larger number of solvents identified by Method II is a probable outcome because the NER model used in Method I has limitations due to its dependency on the training dataset. On the contrary, LLMs leverage contextual understanding along with the brief descriptions provided with the prompts to better identify solvents. Additionally, the LLM can fuse information from different sections of a paper, while Method I relies on paragraph-level segmentation and extraction, which may miss solvents mentioned across different sections or in less explicit contexts. Figure 7 demonstrates an example of how our proposed method can fuse data from different parts of a paper, as given in Ref[55]. Information on chemistries related to perovskite synthesis, such as such as solvents, cations, and synthesis methods, is scattered throughout various sections of the paper. The paragraph on the right represents comprehensive information about perovskite synthesis, which can be used to identify relevant chemicals and processes. The solvent Toluene appears just once in the whole paper but has been identified by the prompting method, which demonstrates its efficiency in fusing sparse information. In the Supporting Information, we have shown how the outputs are generated using both methods for Ref[55].

The solvents in our list that weren't identified by Method I are – Dichlorobenzene, 2-Methoxyethanol, Ethylenediamine, Ethanethiol, and 1-Methyl-2-pyrrolildinone (commonly known as the NMP solvent). We report a complete list of the solvents in the Supporting Information. Dimethylformamide (DMF) is the most frequently used organic solvent, while Ethanol, Dimethylsulfoxide (DMSO), Toluene, and Oleic acid constitute the top five list. DMF is commonly used for the dissolution of lead and Methylammonium (MA) salts[10,56], and hence, it's no surprise that it appears at the top of the list. The distribution of the solvents is similar over different groups of literature articles we downloaded. Since articles were downloaded from different sources, the consistency of distribution of the solvents in these sub-groups is a good reflection of how they are being used in perovskite synthesis. The frequency of appearance of each solvent over different collections of journal articles is given in the supporting information.

Next, we also identified all the organic perovskites mentioned in the synthesis paragraphs we extracted. We were able to acquire more than 350 uniquely mentioned organic perovskites, most of which are MA-based (>40%), while Formamidinium (FA) and Butylammonium (BA) based perovskites constitute around 10% each. A complete list of these perovskites is given in the supporting information. As solvents are required for different activities during perovskite synthesis[12,57], we looked up their mutual distribution in the synthesis paragraphs (see Figure 8). Our study reveals that most of the solvents are reported in conjunction with MA lead halide perovskites. This is not surprising given that the MA-based perovskites have been attractive due to higher efficiency and better stability[58,59]. We further looked into the distribution of these organic perovskites based on their frequency of mutual occurrences with the solvents and plotted the chart shown in Figure 8. This chart shows that out of all the associations between organic perovskites and solvents, more than 3/4$^{th}$ involve MA lead halide perovskites. This reflects the scale of study conducted on these perovskites so far. FA and BA-based perovskites seem to offer alternative choices, but their number is dwarfed by that of the MA-based ones.

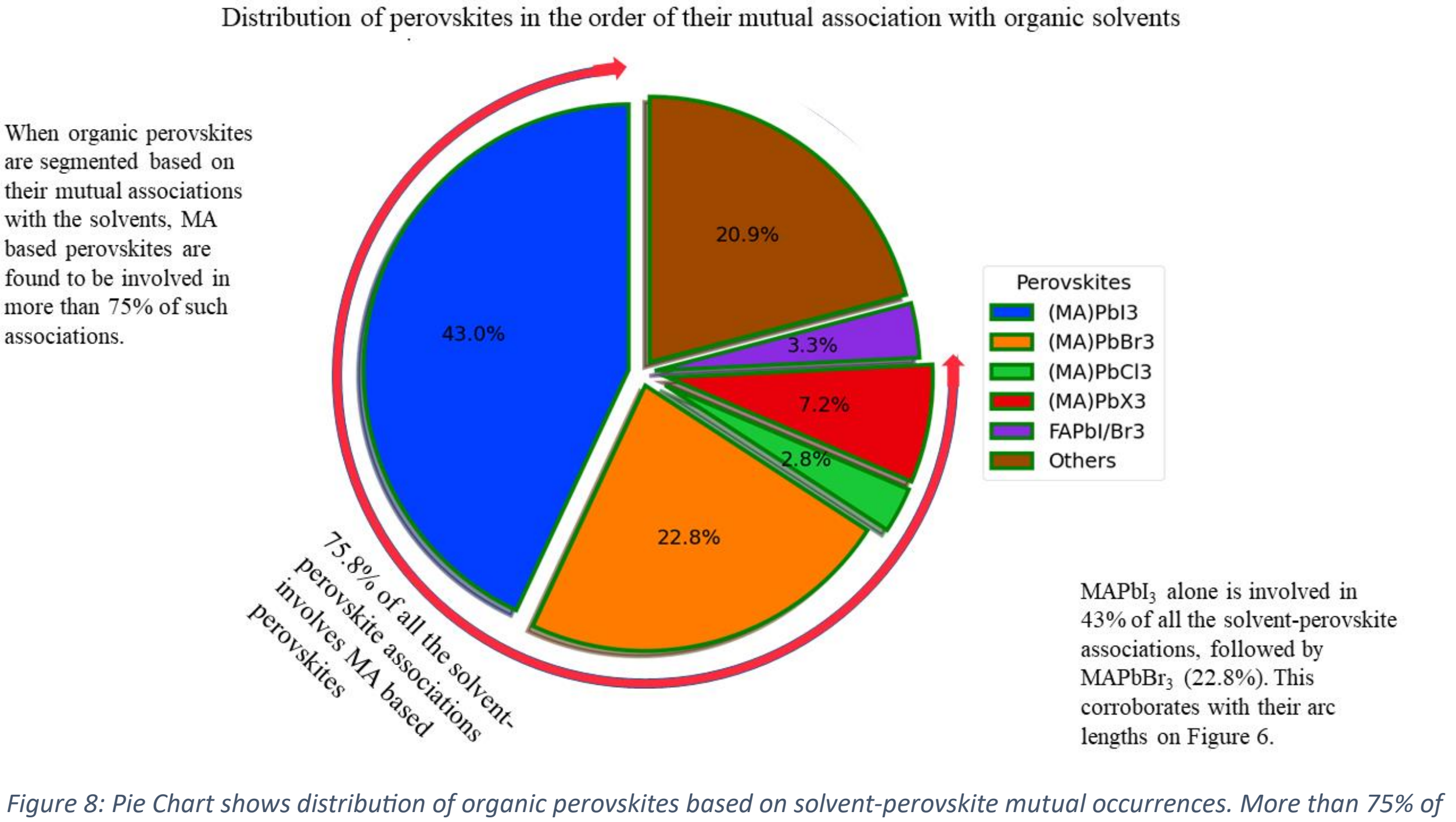

*Figure 8: Pie Chart shows distribution of organic perovskites based on solvent-perovskite mutual occurrences. More than 75% of solvent-perovskite association found in the literature is regarding the methylammonium (MA) lead based perovskites. 26 out of the 36 solvents in our list have associations with these perovskites. A full table of association between the organic perovskites and the solvents is given in the supplementary material.*

These associations are also indicators of the preference priorities of the solvents during perovskite synthesis. More than 70% of the solvents in our list are found to be mentioned in the synthesis paragraphs containing MAPbI3 perovskites. Although these solvents are also used in the synthesis of other perovskites, the majority of them have maximum associations with MAPbI3 only. Solvents such as Hexane, Oleylamine, Octadecene, Oleic acid, Butanol, and Dichloromethane are observed to be mainly used in the synthesis of MAPbBr3 perovskites. Other than MA-based perovskites, Toluene, Isopropanol, and Oleylamine appear more with FA-based perovskites, while DMF, DMSO, and Toluene exhibit greater combinations with BA-based perovskites. This gives an idea of the choice of solvents in the synthesis of specific perovskites

### 3.1 Chord Diagrams Visualizing Mutual Association of Perovskites and Solvents

A chord diagram has been plotted in Figure 9 to further visualize the mutual relation between the most frequently reported perovskites and solvents. A chord diagram is a data visualization tool that represents connections between different entities and is useful in mapping out multivariate associations. In the context of this work, the diagram serves as a crucial tool to screen solvents for further investigation by displaying how frequently identified solvents and perovskites are observed together during perovskite synthesis. The data for a chord diagram is in the form of a matrix, such that the rows and columns have names of the desired entities (in our case, solvents and perovskites), and the values in the matrix determine the width of the chords. This data is given in the supplementary material.

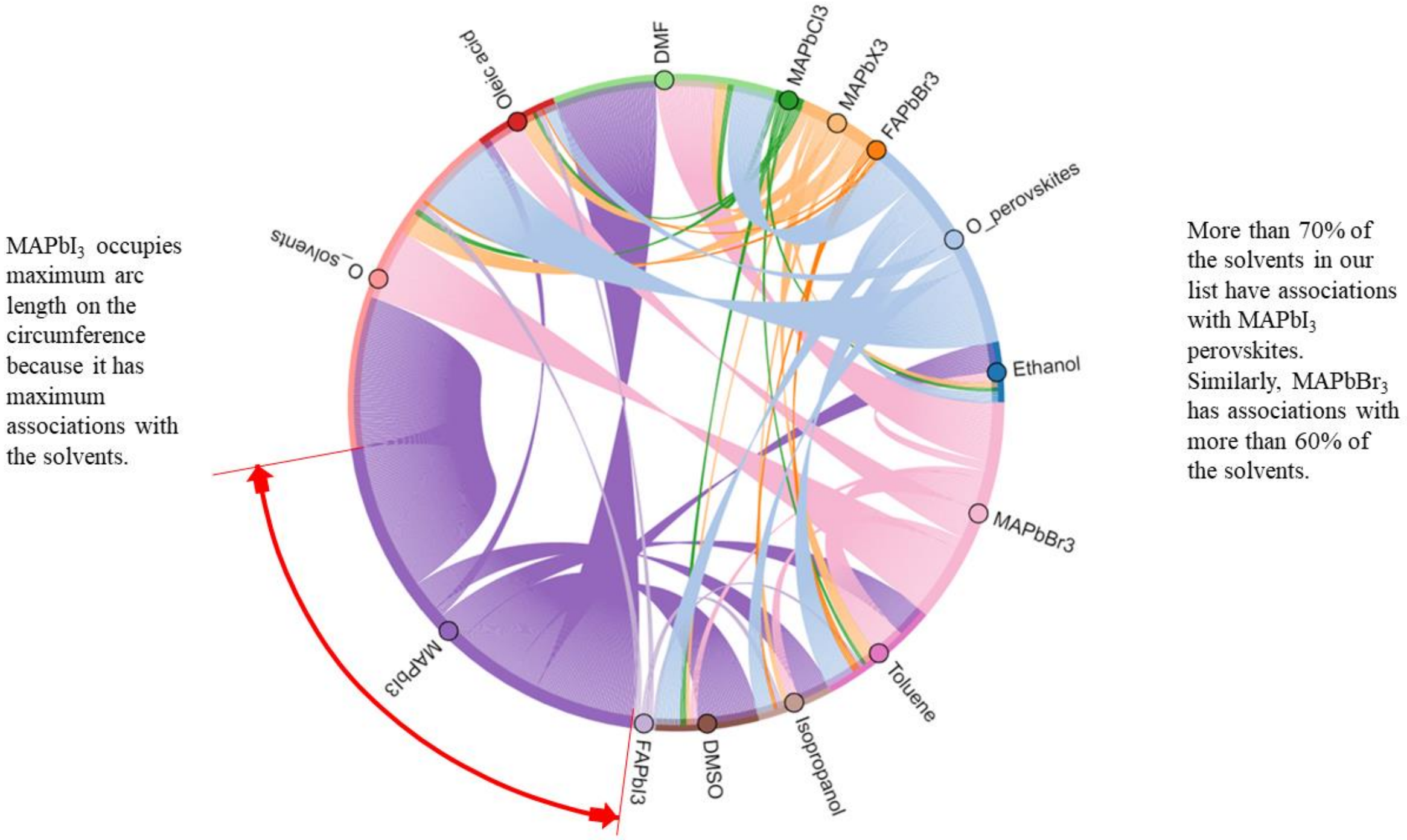

*Figure 9: A chord diagram revealing the associations between highly reported perovskites and the solvents used in their synthesis. O_perovskites and O_solvents represent all other perovskites and solvents, respectively, in our list. Each perovskite and solvent are represented by different colors in different segments of the circumference. The length of the segment is proportional to its frequency and the width of the chords connecting these segments is an indicator of the pair's mutual occurrence. Tracing the chords that link DMF and DMSO with perovskites, one can easily figure out that both these solvents are used in $MAPbI_3$ synthesis while DMF is also highly used for $MAPbBr_3$ synthesis. Similarly, other relationships can be identified.*

The perovskites and solvents are represented by nodes along the circumference of the chord diagram. Each node sits on an arc length of the same color. The length of each arc is proportional to the frequency of occurrence of the perovskite/solvent. The ribbon-like structures (chords) connecting the two arcs are indicators of the frequency of mutual occurrences of the involved nodes. Each chord, although it appears as a single unit, is actually a sum of individual lines connecting the two concerned nodes. A chord made up of n number of lines implies n connections between the nodes. The chord diagram is constructed such that the chords emanate from perovskites and end up in solvents.

For example, $MAPbI_3$ is associated with purple color in the figure, which means it is represented by the purple node, and its connections are represented by the purple chords. The purple arc acts as a base for these purple chords to emerge from and connect to the solvents. The length of this purple arc seems to be the biggest of all, which reflects the maximum mention of $MAPbI_3$ in the literature. The purple chords coming out from $MAPbI_3$ end up in all the solvents. This means $MAPbI_3$ has been synthesized using all these solvents, and their frequency of usage is determined by the thickness of the corresponding chords. Since the chord connecting DMF appears to have a maximum thickness, one can safely assume that DMF is the most used solvent for $MAPbI_3$. One can also gain an insight into the usage distribution of these solvents by examining the arc length occupied by these purple chords. For example, the chords occupy less than 50% of the arc lengths of DMF and Isopropanol, while they occupy 50% or more in the case of Ethanol and DMSO (refer to Figure 9). This means, compared to DMF and Isopropanol, the percentage usage of Ethanol and DMSO is more on the synthesis of $MAPbI_3$, while the higher associations of $MAPbI_3$ with DMF could simply be a result of wider use of DMF. To generalize, this implies that although two chords may have the same thickness, they may not carry the same significance for the concerned association. The thickness of the purple chord is comparable for Ethanol, Isopropanol, and Toluene, but the proportion of arc length they occupy is quite different, thereby reflecting how often that solvent is used in the synthesis of $MAPbI_3$. Solvents are screened for further investigation based on their mutual dependence on perovskites, prioritizing those that frequently appear with common perovskites, such as $MAPbI_3$ and $MAPbBr_3$, rather than just having high individual frequencies

## Classification with Uncertainty

*Table 2: Frequently used organic solvents in perovskite synthesis are categorized into two subclasses (agonist and binding) of active/ inactive endocrine disruptors (EDs). These two subclasses denote a molecule's ability to interact with the estrogen receptor[60]. For a chemical, the state of being active or inactive in one of the subclasses is independent of its nature in the other subclass. However, if the chemical is "Active" in any of the subclasses, then it's potentially an EDC. This classification is done with the help of a deep-learning model that takes SMILES as the inputs and gives a multi-output binary classification. The studies that back up our data for this classification are mentioned in the last column.*

| Index | Solvents | SMILES | ED subclasses | | Reference |
|---|---|---|---|---|---|
| | | | Agonist | Binding | |
| 1 | Dimethylformamide (DMF) | CN(C)C=O | Active | Active | Ref[61,62] |
| 2 | Dimethysulfoxide (DMSO) | CS(=O)C | Inactive | Inactive | |
| 3 | Toluene | CC1=CC==CC=C1 | Active | Active | Ref[63,64] |
| 4 | Oleic acid (OA) | CCCCCCCCC=CCCCCCCCC(=O)O | Inactive | Inactive | |
| 5 | Oleylamine (OLA) | CCCCCCCCC=CCCCCCCCCN | Inactive | Inactive | |
| 6 | Octadecene (ODE) | CCCCCCCCCCCCCCCCC=C | Inactive | Inactive | |
| 7 | Acetone | CC(=O)C | Inactive | Inactive | Ref[65,66] |
| 8 | Chloroform | C(Cl)(Cl)Cl | Inactive | Inactive | |
| 9 | Benzene | C1=CC=CC=C1 | Active | Inactive | Ref[65,67] |
| 10 | Chlorobenzene (CB) | C1=CC=C(C=C1)Cl | Active | Inactive | Ref[68] |
| 11 | Dichlorobenzene (DCB)* | C1=CC(=CC=C1Cl)Cl | Active | Active | Ref[68] |
| 12 | Isopropanol (IPA) | CC(C)O | Inactive | Inactive | |
| 13 | Ethanol | CCO | Inactive | Inactive | |
| 14 | 1-butanol | CCCCO | Active | Active | |
| 15 | 2-Methoxyethanol | COCCO | Active | Active | |
| 16 | Benzyl alcohol | C1=CC=C(C=C1)CO | Inactive | Inactive | |
| 17 | Ethylenediamine (EDA) | C(CN)N | Inactive | Inactive | |
| 18 | Acetonitrile | CC#N | Inactive | Inactive | |
| 19 | n-Hexane | CCCCCC | Inactive | Inactive | Ref[69,70] |
| 20 | Cyclohexane | C1CCCCC1 | Inactive | Inactive | |
| 21 | Diethyl ether | CCOCC | Active | Active | |
| 22 | Dimethyl ether | COC | Inactive | Inactive | Ref[71] |
| 23 | γ – Butyrolactone (GBL) | C1CC(=O)OC1 | Inactive | Inactive | |
| 24 | Methyl acetate | CC(=O)OC | Active | Active | |
| 25 | Ethyl acetate | CCOC(=O)C | Inactive | Inactive | |
| 26 | Ethanethiol | CCS | Inactive | Inactive | |
| 27 | Ethylene glycol | C(CO)O | Inactive | Inactive | |
| 28 | Dichloromethane | C(Cl)Cl | Inactive | Inactive | |
| 29 | n-Octane | CCCCCCCC | Active | Active | Ref[72] |
| 30 | Pyridine | C1=CC=NC=C1 | Inactive | Inactive | |
| 31 | Diethylene glycol (DEG) | C(COCCO)O | Inactive | Inactive | |
| 32 | Sodium hypochlorite | [O-]Cl.[Na+] | Active | Active | |
| 33 | Tetrahydrofuran | C1CCOC1 | Inactive | Inactive | |
| 34 | Trioctylphosphine oxide | CCCCCCCCP(=O)(CCCCCCCC)CCCCCCCC | Inactive | Inactive | |
| 35 | 1-Methyl-2-pyrrolildinone | CN1CCCC1=O | Inactive | Inactive | |

In our analysis, we also categorize frequently used organic solvents in perovskite synthesis, obtained from the chord diagram, into two subclasses of endocrine disruptors (EDs)—'Agonist' and 'Binding'—as shown in Table 2. We have used the deep learning model discussed in Section 3.4 to make our prediction. The studies that substantiate our data are cited in the last column of the table, reinforcing the reliability of our classifications. For example, DMF is listed as a potential endocrine disruptor in a study of chemicals used in natural gas extraction[61]. In a study conducted on workers exposed to DMF in the synthetic leather industry, it has been found to have adverse effects on sperm function[62]. A European analysis of birth weight and length of gestation due to occupational exposure to endocrine disrupting chemicals has listed Toluene as an endocrine disrupting solvent[63]. Such a nature of Toluene has also been established in research that studied low-dose effects and nonmonotonic dose responses of hormones and endocrine disrupting chemicals[64]. Alterations in enzyme activities were reported in rat liver due to n-Octane administration[72]. The endocrine disrupting effect of Benzene has been observed in animals[67] and reported on environmental studies[65]. While these studies reinforce our classifications, there are some conflicting reports as well. Our classification of Acetone as an inactive endocrine disrupting solvent is confirmed in the EPA's report[66], but we also came across an article that says the opposite[65]. Similarly, n-Hexane was reported as a potential EDC in one study[69] but was ruled out in the other[70]. Simply put, for some solvents in our study, there is data to back up their screening as EDC, while for some, there is vague information in the literature, and for the rest, the information is hard to find. However, using a deep learning model that has 90% accuracy, we have given a tool to the scientific community to screen out the potential EDCs when we do not have relevant data on the chemicals. That means our work puts a red flag on these chemicals so that careful consideration is given before using them. In other words, our work can act as a guide in safer solvent selection for perovskite synthesis. For example, almost all solvents have been used in the synthesis of MA lead halide perovskites, but by using this work, one can easily opt for a solvent that is not an active EDC. Both DMF and DMSO are polar solvents and are excellent at dissolving perovskite precursors. However, DMF is an EDC chemical, while DMSO is not. Hence, one can immediately choose to substitute DMF for DMSO in the synthesis of MA lead halide perovskites. Solvents such as Toluene, Isopropanol, and Chlorobenzene are anti-solvents and are used to wash/rinse the solvents to get precursor precipitates[73]. However, Toluene and Chlorobenzene are active EDCs and, hence, are advised to be replaced by Isopropanol or some other anti-solvents with matching properties.

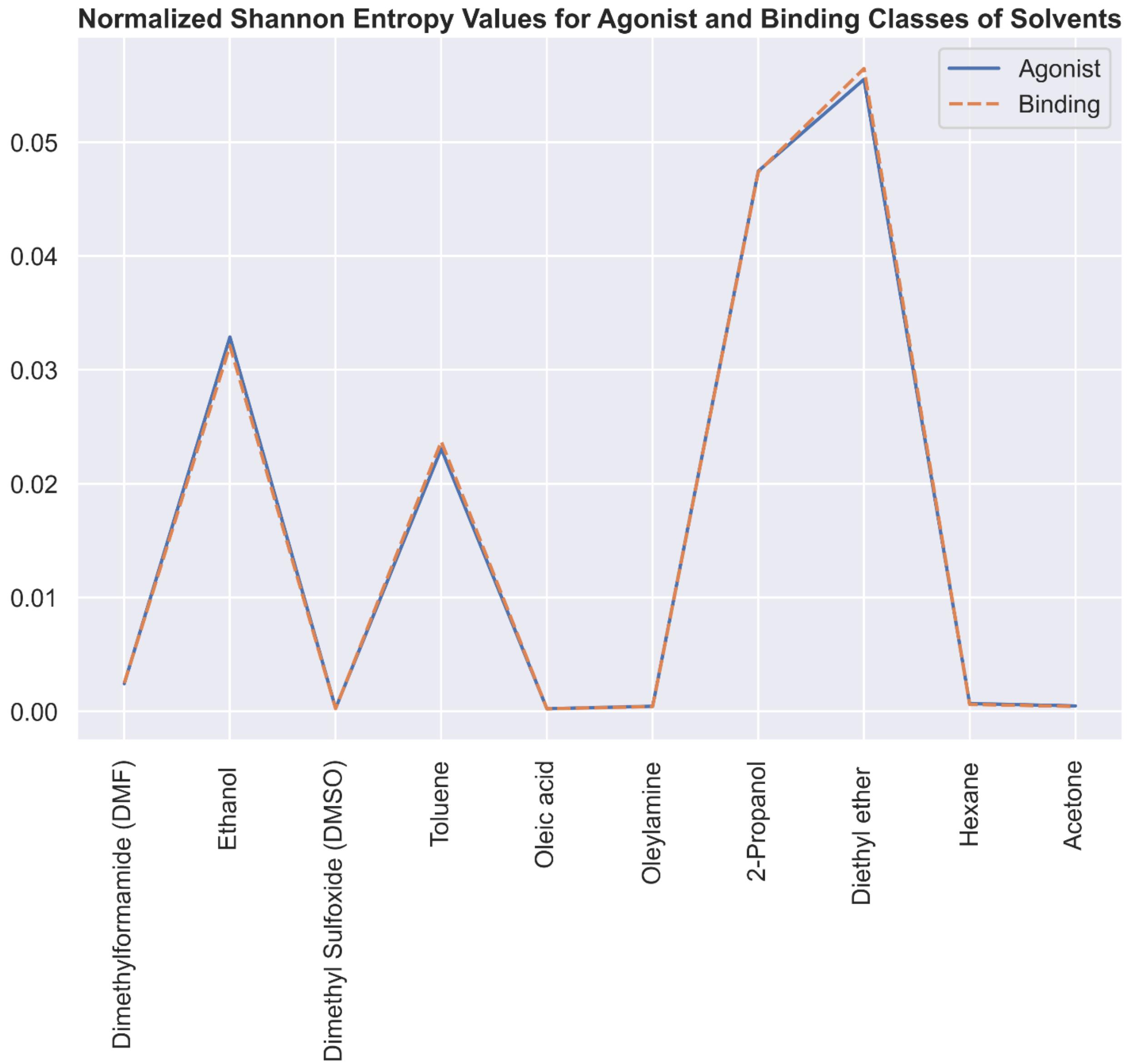

*Figure 10: Uncertainties associated with the prediction of the solvents into agonist and binding classes calculated using Shannon Entropy. A lower value of uncertainty indicates higher confidence in the corresponding prediction. Higher entropy values indicate greater uncertainty in the classification, emphasizing the need for careful consideration and further validation of these results. The orange and blue lines, representing 'Agonist' and 'Binding' classes, respectively, have overlapped, indicating similar levels*

Figure 10 shows the uncertainty computed using the Shannon entropy formula for the ten most frequently appearing solvents used in the synthesis of common perovskites. The figure shows overlapping lines for normalized Shannon entropy values of 'Agonist' (orange) and 'Binding' (blue) classes, indicating similar uncertainty levels in the classification of the solvents. From the figure, Isopropanol and Diethyl ether exhibit higher entropy values, suggesting a lower degree of confidence in their classification, while DMF, DMSO, Oleic acid, Oleylamine, Hexane and Acetone indicate a more confident classification. Our classification model, as explained before, which uses SMILES notation as input, processes these representations through convolutional layers followed by LSTM layers and fully connected layers. As mentioned earlier, the convolution layers extract spatially correlated local features or critical substructures within the molecule, and the LSTM layer maps the sequential dependencies or the order and arrangement of atoms and substructures identified by the convolution layers. Thus, high uncertainty for certain solvents, such as Diethyl Ether and Toluene, may indicate that the chemical substructures within the molecule and their arrangements are difficult for our classification model to identify. The specific structure and/or the substructure may not be well represented in the training dataset.

## 4. Conclusion

In this paper, we have developed an approach for uncertainty-informed toxicity prediction in perovskite synthesis to construct structured data for targeted application from scientific literature and offer reliable toxicity predictions with confidence levels. This is achieved by developing two advanced methods for automated data extraction from scientific literature using contextual language models. These methods focus on gathering pertinent information about organic solvents used in synthesis processes. Method I utilizes smaller, targeted language models such as BERT and ELMo and integrates them with tools such as CDQA and NER to extract relevant data. Method II employs Large Language Models (LLMs) like GPT-3.5, along with novel *prompting and verification* techniques, to ensure the accuracy and reliability of the extracted data. We applied these methodologies to a corpus of 2000 scientific articles on perovskites, enabling the creation of a structured dataset of solvents and perovskites. We have identified 35 solvents using Method I and 54 solvents using Method II.

Our chord diagrams show the prevalence of 70% of solvents such as DMF, DMSO, and Ethanol in synthesizing MAPbI3 and Isopropanol and Oleic acid for FAPbI3. This information is crucial as it highlights the specific solvent-perovskite combinations that optimize device performance and manufacturing efficiency in perovskite-based solar cells. Building on the structured dataset produced by the language models, we employed an uncertainty-informed classification approach using a previously developed deep learning model composed of convolution and LSTM layers. This model uses SMILES representation to assess the endocrine-disrupting potential of solvents. The classification model indicated that 40% of the solvents (e.g., DMF, Toluene) were potential endocrine disruptors. We have used Shannon entropy to quantify the uncertainty of our predictions based on the class probability from the sigmoid layers of the deep neural network, thus providing a measure of confidence in our model outputs and indicating areas that may require further investigation. Results show high confidence for solvents like DMSO and Oleic acid and lower confidence for Toluene and Diethyl ether, requiring further investigation and consideration for expansion of training data.

Expanding our analysis to include a larger corpus and integrating Retrieval-Augmented Generation can further validate the findings from our current work, enhancing our contributions to safer and informed perovskite synthesis practices.